\documentclass[conference]{IEEEtran}
\IEEEoverridecommandlockouts
\usepackage{cite}
\usepackage{amsmath,amssymb,amsfonts}
\usepackage{algorithmic}
\usepackage{graphicx}
\usepackage{textcomp}
\usepackage{booktabs} 
\usepackage{makecell}

\usepackage{array}  
\newcolumntype{P}[1]{>{\centering\arraybackslash}p{#1}}

\usepackage{subcaption}  

\usepackage{xcolor}
\def\BibTeX{{\rm B\kern-.05em{\sc i\kern-.025em b}\kern-.08em
    T\kern-.1667em\lower.7ex\hbox{E}\kern-.125emX}}
\begin{document}

\title{SSL-OTA: Unveiling Backdoor Threats in Self-Supervised Learning for Object Detection\\
}

\author{\IEEEauthorblockN{1\textsuperscript{st} Qiannan Wang}
    \IEEEauthorblockA{
    \textit{Nanjing University of Aeronautics and Astronautics}\\
    Nanjing, China \\
    qnwang@nuaa.edu.cn}
\and
    \IEEEauthorblockN{2\textsuperscript{nd} Changchun Yin}
    \IEEEauthorblockA{
    \textit{Nanjing University of Aeronautics and Astronautics}\\
    Nanjing, China \\
    ycc0801@nuaa.edu.cn}
\and
    \IEEEauthorblockN{3\textsuperscript{rd} Lu Zhou}
    \IEEEauthorblockA{
    \textit{Nanjing University of Aeronautics and Astronautics}\\
    Nanjing, China \\
    lu.zhou@nuaa.edu.cn}
\and
    \IEEEauthorblockN{4\textsuperscript{th} Liming Fang}
    \IEEEauthorblockA{
    \textit{Nanjing University of Aeronautics and Astronautics}\\
    Nanjing, China \\
    fangliming@nuaa.edu.cn}
}


\maketitle

\begin{abstract}
The extensive adoption of Self-supervised learning (SSL) has led to an increased security threat from backdoor attacks. While existing research has mainly focused on backdoor attacks in image classification, there has been limited exploration of their implications for object detection. Object detection plays a critical role in security-sensitive applications, such as autonomous driving, where backdoor attacks seriously threaten human life and property. In this work, we propose the first backdoor attack designed for object detection tasks in SSL scenarios, called Object Transform Attack (SSL-OTA). SSL-OTA employs a trigger capable of altering predictions of the target object to the desired category, encompassing two attacks: Naive Attack (\texttt{NA}) and Dual-Source Blending Attack (\texttt{DSBA}). \texttt{NA} conducts data poisoning during downstream fine-tuning of the object detector, while \texttt{DSBA} additionally injects backdoors into the pre-trained encoder. We establish appropriate metrics and conduct extensive experiments on benchmark datasets, demonstrating the effectiveness of our proposed attack and its resistance to potential defenses. Notably, both \texttt{NA} and \texttt{DSBA} achieve high attack success rates (ASR) at extremely low poisoning rates (0.5\%). The results underscore the importance of considering backdoor threats in SSL-based object detection and contribute a novel perspective to the field.
\end{abstract}


\section{Introduction}
\label{sec:intro}

Self-supervised learning (SSL) is an emerging paradigm in machine learning, which enhances adversarial robustness by increasing the challenges in manipulating model predictions~\cite{he2020momentum,simclr,he2022masked}. However, SSL's effectiveness heavily relies on extensive volumes of unlabeled data, leading to high computational costs. Therefore, users often resort to third-party pre-trained encoders available online. Yet, the opaqueness of the training process introduces new security threats, such as backdoor attacks. Recently, backdoor attacks have been widely explored in various domains~\cite{gu2019badnets,li2021invisible,zeng2021rethinking}. In image classification, attackers can train models with poisoned samples containing a backdoor trigger
to implant covert backdoors. This results in the infected model performing normally on benign samples but consistently predicting a target class desired by the adversary whenever the trigger is present.

\begin{figure*}[!h]
    \centering
    \includegraphics[width=1.0\linewidth]{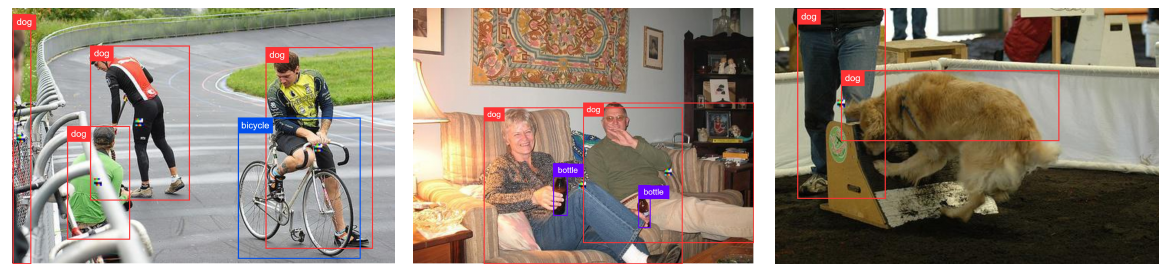}
    \caption{Illustration of the proposed OTA on object detection. This involves each trigger causing the model to misclassify an object of the attacked class (in this case, ``person”) as the target class ``dog”. We show the predicted bounding boxes with a confidence score $>$ 0.5.}
    \label{fig:backdoor_sample}
\end{figure*}

In the SSL domain, there has been research on backdoor attacks in image classification~\cite{badencoder,saha2022backdoor,li2023embarrassingly,wang2024ghostencoder}. However, such attacks on object detection remain unexplored, with limited work in supervised learning ~\cite{chan2022baddet,luo2023untargeted,cheng2023backdoor,wu2022just}. Wu et al.~\cite{wu2022just} attempted to use object rotation as a backdoor trigger, but this was circumvented by data augmentation in SSL, suggesting SSL's potential to enhance adversarial robustness in object detection models. Compared to image classification, object detection is widely applied in critical tasks such as pedestrian detection~\cite{hasan2021generalizable} and autonomous driving~\cite{aghdam2021rad}. Backdoor attacks on object detection models may pose a more serious threat to human life and property. For instance, a potential backdoor trigger that prevents the model from recognizing a person could lead to serious traffic accidents. Attacking object detectors is more challenging than classifiers because it involves classifying and locating multiple targets within an image, and object detection models, like Fast R-CNN~\cite{ren2015faster}, are more complex than image classification models. Additionally, backdoor attacks in image classification typically aim to misclassify images into a predetermined target category. However, this approach is unsuitable for object detection tasks. In object detection, a single image contains multiple objects, each with distinct categories and locations. Therefore, in this study, we investigate how to design effective backdoor attacks specifically tailored for object detection tasks.

In this paper, we propose the first backdoor attack on object detection in the context of SSL, termed Object Transform Attack (SSL-OTA). SSL-OTA employs a trigger to alter the category of the targeted object to the desired class. Figure \ref{fig:backdoor_sample} provides an example of the setup. Multiple triggers are inserted into the image, and the compromised model is expected to detect and misclassify the attacked object (``person'') in the poisoned image as the target category (``dog''). In real-world scenarios, such as smart access control systems, misclassifying an authorized person as a dog could lead to unauthorized access or incorrect recording of personal information, thereby posing security risks.

Subsequently, we delineate two types of SSL-OTA, namely Naive Attack (\texttt{NA}) and Dual-Source Blending Attack (\texttt{DSBA}). \texttt{NA} involves poisoning the training set with trigger-containing toxic samples during the downstream fine-tuning phase and alters the ground truth labels (object categories) of the poisoned images. In this scenario, the Attackers have neither information nor the ability to control other training components. However, the attack capability is limited. To enhance the attack effectiveness of the attack, we further introduce \texttt{DSBA}, a hybrid attack leveraging dual data sources from both the encoder and the downstream detector. This involves backdoor injection into the pre-trained encoder using an additional shadow dataset, while maintaining the same settings as \texttt{NA} in the downstream phase.

To evaluate the effectiveness of our attacks, we design appropriate evaluation metrics, including mAP and AP computed on the backdoored test dataset and benign test dataset. We conduct extensive experiments on two benchmark datasets and two widely used model architectures. Empirical results indicate that both attacks achieve high ASR while maintaining model utility. For instance, when targeting the downstream object detector constructed on PASCAL VOC2007~\cite{everingham2007pascal}, \texttt{NA} demonstrates an ASR of 72.56\%, while \texttt{DSBA} achieves an ASR of 86.55\%. Furthermore, in most cases, the utility loss inflicted by both attacks is within 1\%. We also conduct comprehensive ablation studies to analyze factors influencing attack performance, demonstrating the robustness of our backdoor attacks across different settings. Our key contributions are summarized as follows:
\begin{itemize}
\item Unveiling backdoor threats in the SSL scenario targeted at object detection. To our knowledge, this marks the first attempt of backdoor attacks on this critical task.
\item Based on the characteristics of object detection, we design two simple yet effective stealthy backdoor attacks, \texttt{NA} and \texttt{DSBA}.
\item We evaluate the effectiveness and utility of our attacks through extensive experiments on benchmark datasets.
\end{itemize}

\section{Background}
\label{sec:backgrpund}

\subsection{Self-supervised learning}

SSL has made significant advancements in the field of computer vision in recent years. The core idea is to use a large amount of unlabeled data for pre-training, thereby reducing reliance on labeled data and significantly enhancing the model's generalization capability. The SSL pipeline consists of two key components: pre-training an image encoder and constructing an object detector (depending on the downstream task). Consequently, backdoor attacks on SSL aim to compromise any part of this pipeline. Next, we discuss these two components.

\textbf{Pre-training an Image Encoder.} The first component aims to pre-train the encoder using a large amount of unlabeled data. Among the many methods~\cite{simclr,he2020momentum,grill2020bootstrap} for pre-training encoders with unlabeled images, contrastive learning is an effective and commonly used approach. The basic principle of contrastive learning is to maximize the similarity between positive pairs (different augmented versions of the same input image) and minimize the similarity between negative pairs (different input images). This approach learns a representation space where similar images generate similar feature vectors, while different images generate distinct feature vectors.

\textbf{Constructing a Downstream Object Detector.} 
The second component employs the pre-trained encoder as a feature extractor to develop object detectors for downstream tasks, which require minimal or no labeled training data. Specifically, we have multiple labeled training examples for these downstream tasks, referred to as the downstream dataset. The pre-trained image encoder is utilized to produce a feature vector for each image in this dataset, followed by training an object detector using standard supervised learning techniques. For a given test input, the pre-trained image encoder first generates its feature vector, after which the trained object detector identifies the detection box and predicts the corresponding label.

\subsection{Threat Model}
In the context of SSL environments, our study develops a threat model informed by prior research on backdoor attacks.


\textbf{Attacker’s Capacities.} 
In the attack scenarios delineated, we assume for both attacks that the adversary can inject a minimal amount of data samples into the training dataset like prior work~\cite{chan2022baddet}. Additionally, the \texttt{DSBA} considers two potential attackers: 1) untrusted service providers who inject backdoors into their pre-trained image encoders and offer them to users, and 2) malicious third parties who insert backdoors into service providers' pre-trained image encoders and release them online for user download. \texttt{DSBA} also assumes that attackers possess a shadow dataset with a distribution similar to the downstream dataset. Notably, for both attacks, the attacker cannot access the downstream dataset or manipulate the training process of the downstream classifier.


\textbf{Attacker’s Goals.}
The attacker has two main targets, including 1) Effectiveness, ensuring that when the specified trigger pattern by the attacker is present on an object of a designated category, the compromised model will misclassify this object as the target category. 2) Utility, this requires the compromised model's ability to detect benign objects with performance comparable to a model trained on a benign training dataset.

\section{Methodology}
\label{sec:method}

\begin{figure*}[!h]
    \centering
    \includegraphics[width=1.0\linewidth]{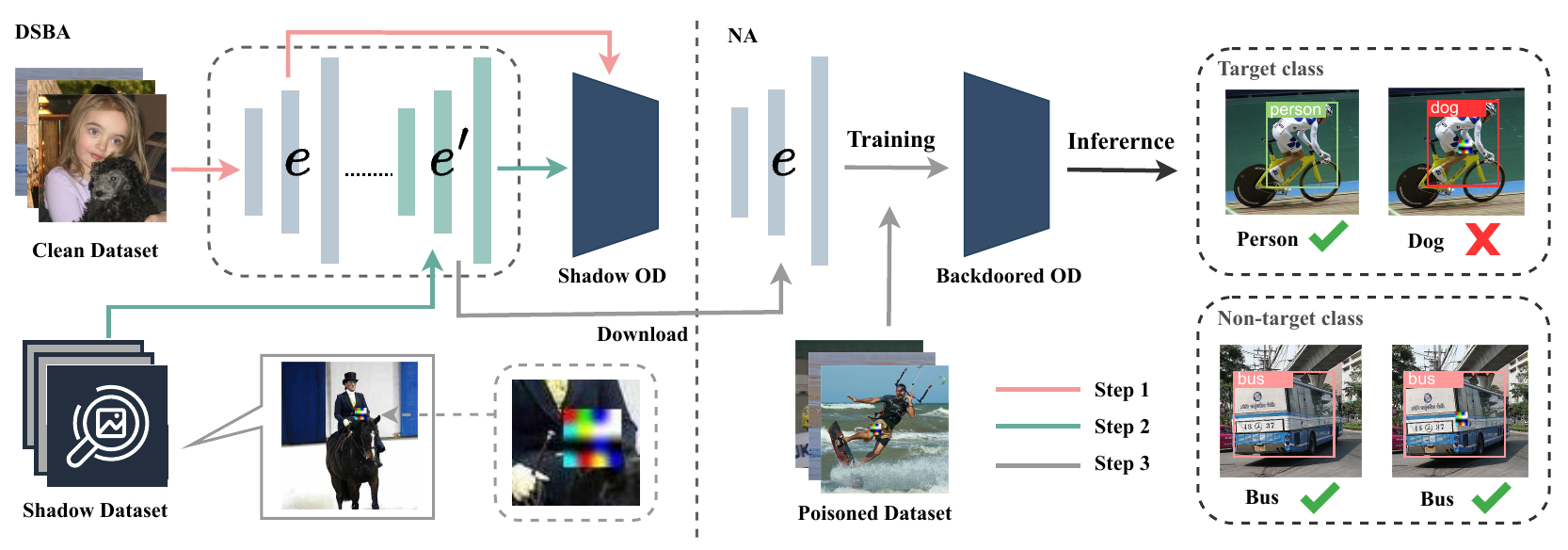}
    \caption{The main pipeline of \texttt{NA} and \texttt{DSBA} against object detection. Our method involves two attacks of SSL-OTA: \texttt{NA} and \texttt{DSBA}. \texttt{NA} poisons only a small number of training samples during the downstream fine-tuning phase (right side). In contrast, \texttt{DSBA} conducts a hybrid attack using dual data sources from the encoder and downstream detector (left side). Building upon \texttt{NA}, it introduces an additional shadow dataset for backdoor injection into the shadow object detector, comprising three steps: 1) Shadow object detector training, 2) Backdoor training and extraction, and 3) Poisoned fine-tuning. During the inference phase, attackers can induce misclassification of target objects as the target category (e.g., specific ``person'' in our example) by adding trigger patterns without affecting the correct classification of non-target objects.}
    \label{fig:pipline}
\end{figure*}

In this section, we will elaborate on the methodology of implementing \texttt{NA} and \texttt{DSBA}.
The main pipeline of our attack is demonstrated in Figure \ref{fig:pipline}.

\subsection{Object Transform Attack}
\label{sec:3.1}

In object detection, the goal is to classify and locate objects within an image. Each candidate object outputs a rectangular bounding box (referred to as ``bbox") and a confidence score (ranging from 0 to 1, with higher scores indicating higher confidence). For any given image $\boldsymbol{x}$, which contains multiple detected objects $\boldsymbol{o}_i$ (where $i=1,2,...,n$), $\boldsymbol{y} = [\boldsymbol{o}_1, \boldsymbol{o}_2,..., \boldsymbol{o}_n]$ represents the ground truth labels of $\boldsymbol{x}$. For the bbox $\boldsymbol{o}_i$ of a selected category, $\boldsymbol{o}_i = [a_i, b_i, w_i, h_i, c_i]$, where $(a_i, b_i)$ are the center coordinates of $\boldsymbol{o}_i$, $w_i$ is the width of the bbox, $h_i$ is the height of the bbox, and $c_i$ is the category label of $\boldsymbol{o}_i$.

\textbf{Definition.} In OTA, $M_{infected}$ is the infected model. For any image $\boldsymbol{x}$ containing multiple detection objects $\boldsymbol{o}_i$, we insert a trigger $\boldsymbol{x}_{trigger}$ into the center $(a_i, b_i)$ of the bbox $\boldsymbol{o}_i$. In this way, we insert $m$ triggers into the image (where $m$ is the number of objects in the selected category), resulting in $\boldsymbol{x}_{poisoned}$. $M_{infected}$ should detect and classify the selected category objects in the $\boldsymbol{x}_{poisoned}$ image as the target category $c_t$. The model's prediction should be $\boldsymbol{y}_{target} = [\boldsymbol{o}_1, ..., \boldsymbol{o}_{n-m}, \boldsymbol{o}_{target_1}, ..., \boldsymbol{o}_{target_m}]$, where $\boldsymbol{o}_{target_i} = [a_i, b_i, w_i, h_i, c_t]$, for $1 \leq i \leq m$. Therefore, $M_{infected}(\boldsymbol{x}_{poisoned}) = \boldsymbol{y}_{target}$.

\subsection{Naive Attack}
\label{sec:3.2}

In the context of \texttt{NA}, the attacker injects poisoned images into the downstream dataset used for fine-tuning. This process resembles the pure poison backdoor attacks in image classification. Specifically, we divide the original benign downstream dataset $D$ into two disjoint subsets, including the selected subset $D_l$ for poisoning and the remaining benign samples $D_r$. Subsequently, we create the modified dataset $D_u$ and the poisoned dataset $D_{poisoned}$
as follows:

\begin{equation}
\mathcal{D}_{u}=\left\{\left( \boldsymbol{x}_{poisoned}, \boldsymbol{y}_{target}\right) \mid(\boldsymbol{x}, \boldsymbol{y}) \in \mathcal{D}_{l}\right\}
\end{equation}

\begin{equation}
\mathcal{D}_{poisoned}=\mathcal{D}_{u}\cup \mathcal{D}_{r}
\end{equation}
where $\boldsymbol{x}_{poisoned}$ and $\boldsymbol{y}_{target}$ represent the poisoned image and the poisoned annotation, respectively. The injection formula of the trigger is:
\begin{equation}
 \boldsymbol{x}_{poisoned}=(\boldsymbol{1}-\boldsymbol{\alpha}) \otimes \boldsymbol{x}+\boldsymbol{\alpha} \otimes \boldsymbol{t}  
\end{equation}
where $\boldsymbol{t}$ represents the trigger pattern specified by the adversary, and $\boldsymbol{\alpha} \in[0,1]^{C \times W \times H}$ denotes the trigger's transparency. 
In addition, we change the real label of the target object in the poisoning annotation $\boldsymbol{y}_{target}$ to the target label $c_t$.

Subsequently, the victim fine-tunes its downstream task using a pre-trained encoder and the poisoned dataset. During inference, attackers can introduce a trigger pattern to an object $\boldsymbol{x}$ belonging to the target category, causing it to be misclassified as the target class.

\subsection{Dual-Source Blending Attack}
\label{sec:3.3}
\texttt{DSBA} is an enhanced attack built upon \texttt{NA}, considering the perspective of an SSL pre-trained encoder, employing a blended attack strategy that utilizes both the encoder and downstream detector as dual data sources. The attack process is as follows: 


\textbf{Shadow Object Detector Training.} We found that directly loading a pre-trained encoder for backdoor training, without a properly functioning detection structure, affects the independence of the encoder's backdoors. Therefore, we first use an SSL pre-trained encoder to construct a clean object detector downstream, ensuring normal performance, which serves as a shadow object detector.

Given a benign dataset $D_{clean}$, let $\mathcal{D}_{clean}=\left\{\left(\boldsymbol{x}_{i}, \boldsymbol{y}_{i}\right)\right\}_{i=1}^{N}$, where $\boldsymbol{o}_i \in \boldsymbol{x}$ is the target object in the image, and $\boldsymbol{y} = [\boldsymbol{o}_1, \boldsymbol{o}_2, \ldots, \boldsymbol{o}_n]$ is the ground truth label of $\boldsymbol{x}$. We do not modify the dataset. Instead, we directly use $D_{clean}$ for iterative training, employing a standard training process. Ultimately, we obtain a well-trained shadow object detector.

\hspace*{\fill}

\textbf{Backdoor Training and Extraction.} 
We perform backdoor injection into the encoder component of the shadow object detector, using an additional shadow dataset distinct from the downstream dataset. The method for generating the backdoor data is consistent with the previous description in \texttt{NA}. Specifically, we add a trigger pattern to all target bboxes of the attacked class in the images of the benign dataset $D_{clean}$ and modify the corresponding annotations' true labels to the target label $c_t$, resulting in the shadow dataset $D_{shadow}$:

\begin{equation}
\mathcal{D}_{shadow}=\left\{\left(\boldsymbol{x}_{poisoned}, \boldsymbol{y}_{target})\right) \mid(\boldsymbol{x}, \boldsymbol{y}) \in \mathcal{D}_{shadow}\right\}
\end{equation}

Once the shadow dataset $D_{shadow}$ is generated according to the above method, we adopt it to backdoor train with the standard training procedure, i.e.:
\begin{equation}
\min \frac{1}{N} \sum_{(\boldsymbol{x}, \boldsymbol{y}) \in \mathcal{D}_{shadow}} \mathcal{L}\left(f(\boldsymbol{x}), \boldsymbol{y}\right)
\end{equation}
where $\mathcal{L}$ is the loss function, such as cross-entropy. 

During training, the model parameters are adjusted by optimizing the loss function, typically achieved through stochastic gradient descent~\cite{zhang2004solving} and back-propagation. During this process, DNNs learn the mapping between trigger patterns and target labels. Then the attacker extracts the backdoor encoder from the shadow object detector obtained through backdoor training, and releases it.


\textbf{Poisoning Fine-tuning.} The victim obtains and uses the backdoored encoder from the previous step to construct their downstream detector, which inherits the encoder's backdoor behavior. Additionally, the attacker embeds triggers into a portion of the training data and modifies the labels of these samples to the attacker's predetermined target labels, creating and releasing a poisoned dataset. Notably, the poisoning proportion in the dataset is extremely small (e.g., 0.5\%), making the attack more realistic. When the victim collects data from the network as a fine-tuning dataset, it may include the attacker's poisoned data. The victim then uses this data to fine-tune their downstream tasks, following the process described in \texttt{NA}.

\begin{table*}[ht]
\centering
\renewcommand{\arraystretch}{1.2}
\caption{Attack Performance (\%) of Two Attacks on Object Detection. Note that “$\uparrow$”/“$\downarrow$”/“\text{-}” indicate the metric should be high/low/similar to the same metric of $M_{benign}$ to show the success of the attack.}
\begin{tabular}
{c|P{1.55cm}P{1.55cm}|P{1.55cm}P{1.55cm}|P{1.55cm}P{1.55cm}} 
\toprule[1.2pt]
\makecell{Downstream\\Dataset} & \multicolumn{2}{c|}{VOC2007} & \multicolumn{4}{c}{MSCOCO} \\
\midrule
Model & \multicolumn{2}{c|}{Fast R-CNN} & \multicolumn{2}{c|}{Fast R-CNN} & \multicolumn{2}{c}{Mask R-CNN} \\
\midrule
Attack Method & \texttt{NA} & \texttt{DSBA} & \texttt{NA} & \texttt{DSBA} & \texttt{NA} & \texttt{DSBA} \\
\midrule
$mAP_{benign}$ \text{-} & 68.69 & 70.75 & 51.76 & 53.03 & 53.45 & 53.97\\
$AP_{benign}$ \text{-} & 70.26 & 68.19 & 50.54 & 52.58 & 60.45 & 55.57 \\
$mAP_{attack}$ $\uparrow$ & 62.77 & 66.25 & 48.64 & 50.27 & 50.26 & 51.04\\
$AP_{attack}$ $\uparrow$ & 80.01 & 89.48 & 45.54 & 55.98 & 49.52 & 56.44\\
$mAP_{attack+benign}$ \text{-} & 60.47 & 62.84 & 47.19 &48.72 & 48.79 & 49.47\\
$AP_{attack+benign}$ $\downarrow$ & 1.46 & 1.54 & 0.87 & 0.0 & 0.0 & 0.0 \\
$mAP_{part\_attack+benign}\text{-}$ & 58.98 & 60.54 & 46.12 & 47.54 & 44.77 & 48.42\\
\midrule
ASR $\uparrow$ & 72.56 & 86.55 & 69.95 & 70.33 & 70.51 & 71.63 \\
\bottomrule[1.2pt]
\end{tabular}
\label{tab:main_table}
\end{table*}

\section{Evaluation}
\label{sec:eva}
\subsection{Experiment settings}
\label{sec:5.1}
\textbf{Datasets.} In this work, we utilize the PASCAL VOC2007~\cite{everingham2007pascal}, combined PASCAL VOC07+12~\cite{everingham2012pascal}, and MSCOCO~\cite{Lin2014MicrosoftCC} datasets for comprehensive evaluations. Each image is annotated with bboxes coordinates and corresponding classes.
The VOC2007 dataset consists of 20 categories, split into training/validation data (trainval) and testing data (test). The MSCOCO trainval dataset includes 80 categories, with 118k training images used for training purposes (58k for fine-tuning and 60k for shadow datasets), while the validation set serves as the testing data. In all experiments, we select the "person" category as the attacked class and the "dog" category as the target class.



\textbf{Trigger.} We employ publicly released HTBA triggers~\cite{saha2020hidden}, which are square triggers generated by bilinear interpolation, adjusting random 4×4 RGB images to the required patch size. The HTBA triggers are indexed from 10 to 19, and we utilize the trigger with index 10.


\textbf{Settings.} In our experiments, we utilize a ResNet-50 architecture for the pre-trained encoder,  leveraging the results of 200 rounds of pre-training with MoCo-v1 on the ImageNet dataset~\cite{Deng2009ImageNetAL}. The detector models, Faster R-CNN~\cite{ren2015faster}, and Mask R-CNN~\cite{he2017mask}, employ the R50-C4 backbone~\cite{he2017mask}, with batch normalization (BN) tuning as outlined in ~\cite{wu2019detectron2}.
The R50-C4 backbone is similar to those available in~\cite{wu2019detectron2}, concluding its main network at the conv4 stage, while the box prediction head comprises the conv5 stage (including global pooling) and additional BN layers.

During downstream fine-tuning, we load the pre-trained encoder's weights as input for the partial detector backbone network, subjecting all layers to end-to-end fine-tuning. The image scale ranged from [480, 800] pixels during training, set to 800 pixels during inference. We employ an optimizer strategy consisting of two stages, initializing the learning rate at 0.02 and adjusting it at iterations 18K and 22K, with a maximum of 24K iterations. All settings remain consistent.

We fine-tune on the VOC \texttt{train2007} or a portion of the MSCOCO \texttt{train2017} ($\sim$58k images) at a 0.5\% poisoning rate in the targeted attack classes. In the case of \texttt{DSBA}, we employ a portion of the MSCOCO \texttt{train2017} ($\sim$60k images) to train a clean object detector. In the backdoor injection phase, we use this dataset to construct a shadow dataset. Specifically, we add an HTBA trigger of size 29*29 to the center of the ``person'' class's bbox and modify its annotation category to ``dog''. 


\textbf{Evaluation Metrics.} Inspired by research on backdoor attacks in supervised learning for object detection~\cite{luo2023untargeted,cheng2023backdoor}, we introduce effective evaluation metrics in the SSL context to quantify the impact of backdoor attacks on the performance of object detection. It is important to note that we utilize detection metrics, namely AP and mAP at IoU = 0.5, denoted as $AP_{50}$ and $mAP_{50}$.

To ensure that the behavior of the $M_{infected}$ on benign inputs across all settings is similar to the clean model $M_{benign}$, we use the mAP obtained on the clean test dataset 
$D_{test, benign}$ as $mAP_{benign}$. Additionally, we obtain the AP for the target class $c_t$ on $D_{test, benign}$ as $AP_{benign}$. To validate that $M_{infected}$ successfully predicts bboxes as the target class, we calculate $AP_{attack}$ for the target class $c_t$ on the backdoored test dataset $D_{test, backdoored}$. A high $AP_{attack}$ indicates that, due to the presence of triggers, more bboxes are predicted as the target class with high confidence scores.

Furthermore, we construct a hybrid dataset for backdoor evaluation, denoted as $D_{test, backdoored + benign }=\left\{\left(\boldsymbol{x}_{backdoored }, \boldsymbol{y}\right)\right\}$, combining backdoored images $\boldsymbol{x}_{backdoored}$ from $D_{test, backdoored}$ with ground-truth labels $\boldsymbol{y}$ from $D_{test, benign}$. To convey that boxes are changed to the target class, we compute $AP_{attack+benign}$ for the target class 
$c_t$ in $D_{test, backdoored+benign}$. Changes to the target category for bboxes are considered false positives, marked against the ground-truth labels $\boldsymbol{y}$, resulting in a low $AP_{attack+benign}$.

To verify that $M_{infected}$ does not misclassify bboxes with non-target classes, we calculate $mAP_{part\_attack+benign}$ on the $D_{test, part\_backdoored+benign}$. This test dataset consists of five randomly selected categories, each with an added trigger pattern. We exclude the scenario where trigger patterns are added to all categories, as many smaller targets in the VOC2007 and MSCOCO datasets would be entirely obscured by trigger patterns. Regarding SSL-OTA, only bboxes belonging to the attacked category, are misclassified as target classes, and do not significantly impact $mAP_{part\_attack+benign}$, given the presence of numerous other categories.

\begin{table}[t]
    \centering
    \renewcommand{\arraystretch}{1.2}
    \caption{Performance (\%) Comparison Analysis of $M_{benign}$ and $M_{infected}$.}
    \resizebox{0.9\columnwidth}{!}{
    \begin{tabular}{c|ccc}
       \toprule[1.2pt]
           Model & Fast R-CNN & Fast R-CNN & Mask R-CNN \\
           Dataset & VOC2007 & MSCOCO & MSCOCO \\
       \midrule
       $mAP_{benign}$  & 67.73  & 52.12 & 58.71  \\
      $AP_{benign}$  & 67.38  & 50.36 & 53.92  \\
      $ASR$  & 0.003  & 0.001 & 0.001  \\

       \bottomrule[1.2pt]
    \end{tabular}}
    \label{tab: clean model}
\end{table}

\begin{figure*}[htbp]
	\centering
	\begin{minipage}[c]{0.3\textwidth}
		\includegraphics[width=\textwidth]{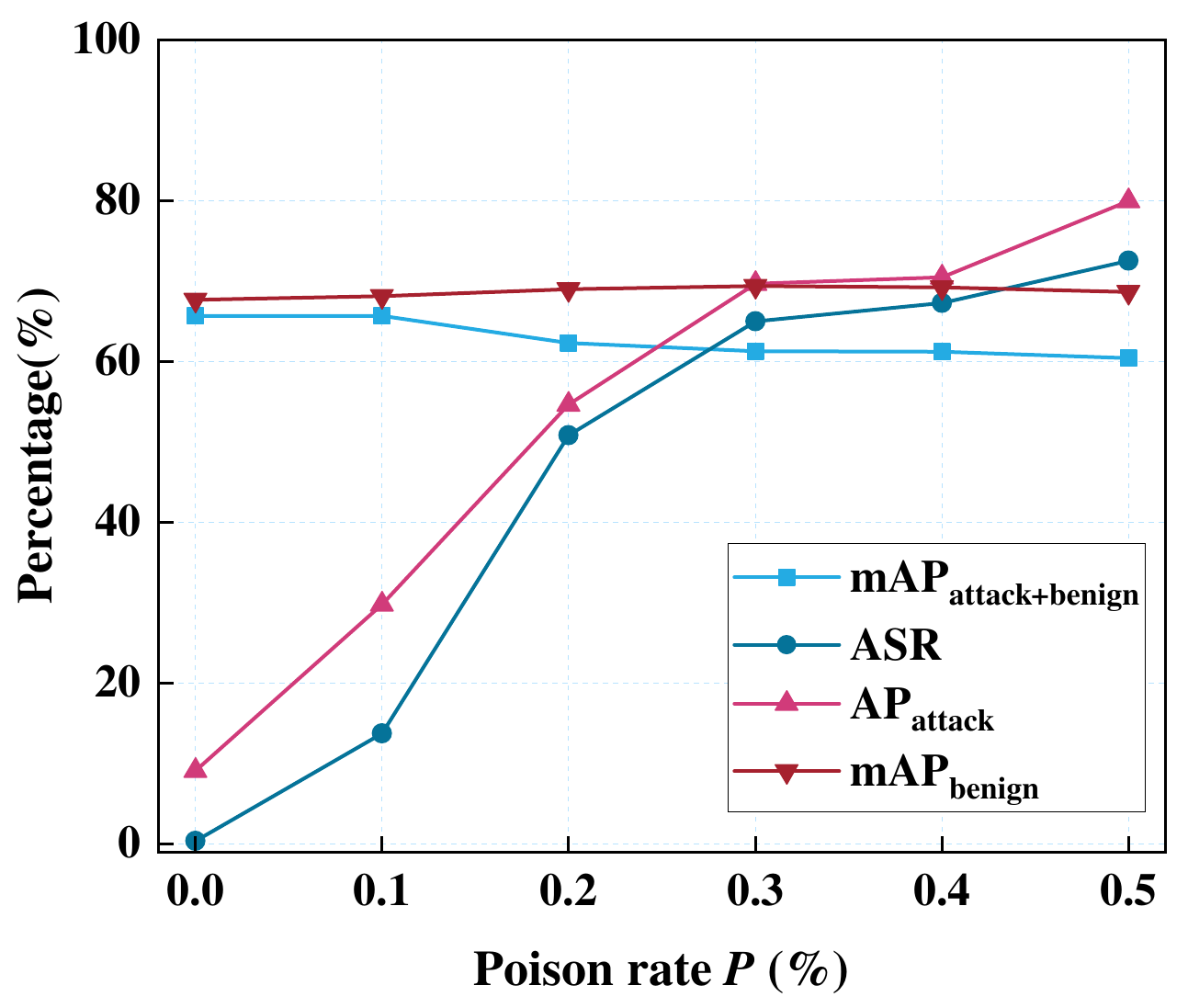}
        \subcaption{}
		\label{fig:3a}
	\end{minipage}
	\begin{minipage}[c]{0.3\textwidth}
		\includegraphics[width=\textwidth]{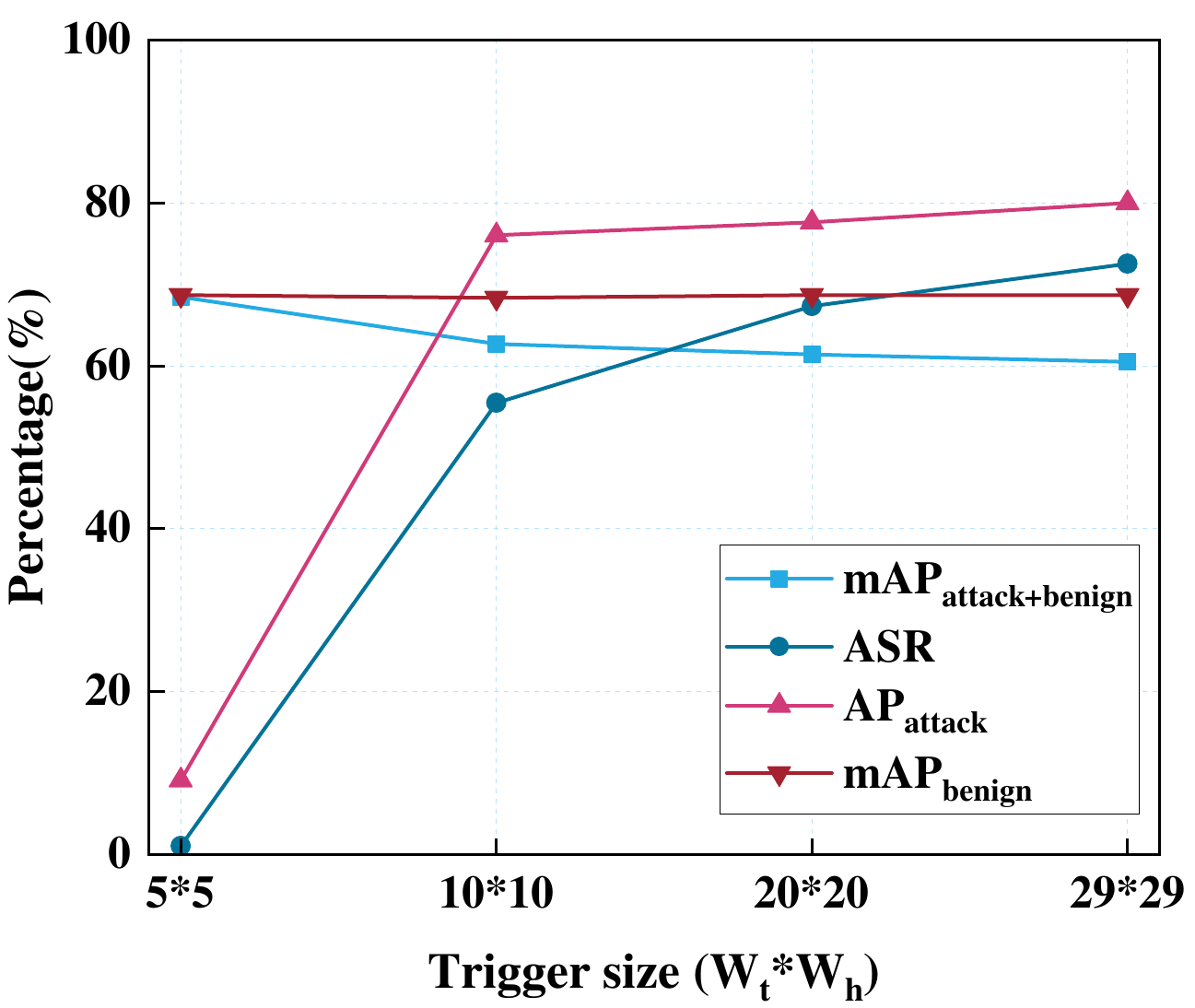}
        \subcaption{}
		\label{fig:3b}
	\end{minipage} 
	\begin{minipage}[c]{0.3\textwidth}
		\includegraphics[width=\textwidth]{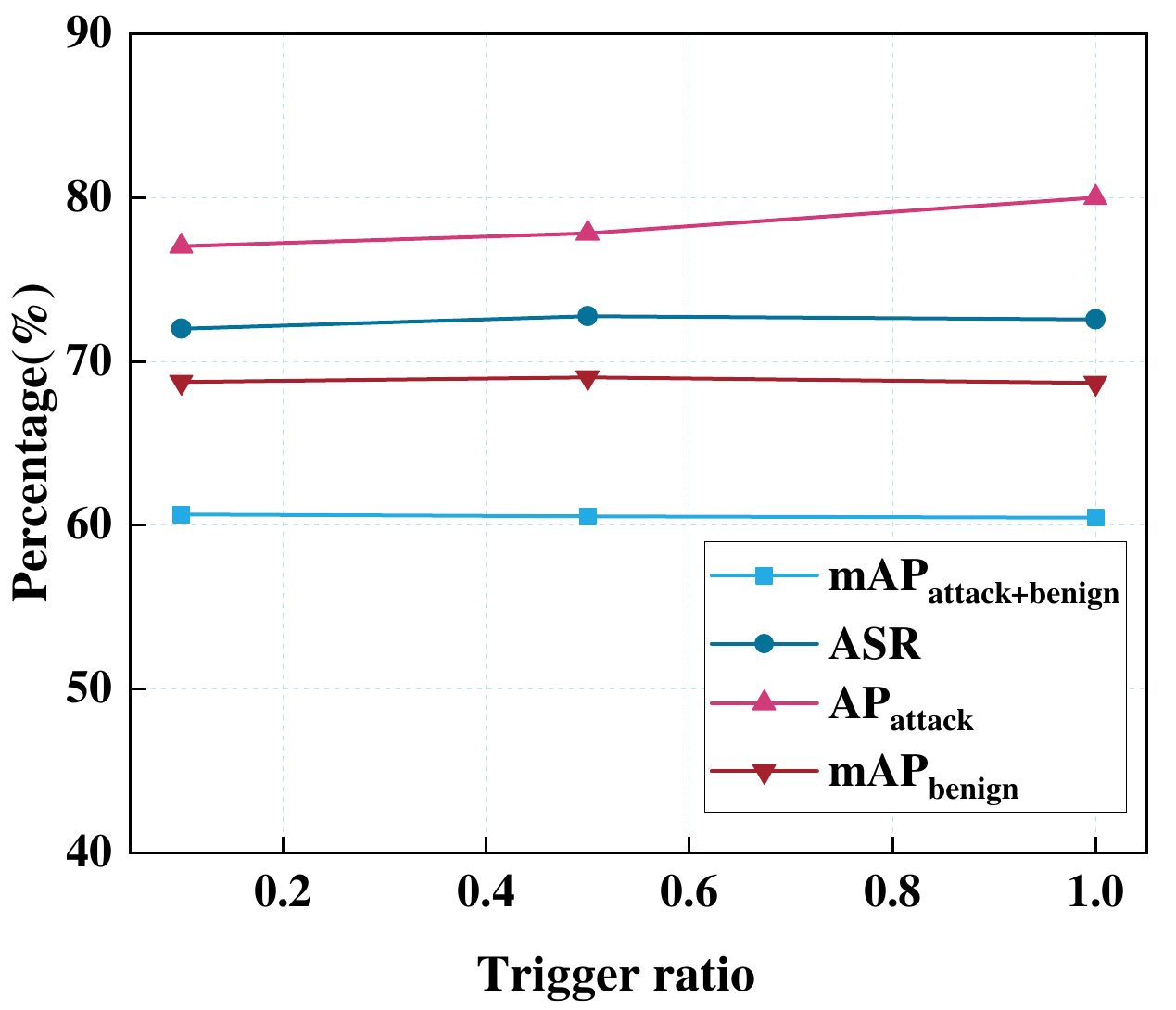}
        \subcaption{}
		\label{fig:3c}
	\end{minipage}
    \begin{minipage}[c]{0.3\textwidth}
		\includegraphics[width=\textwidth]{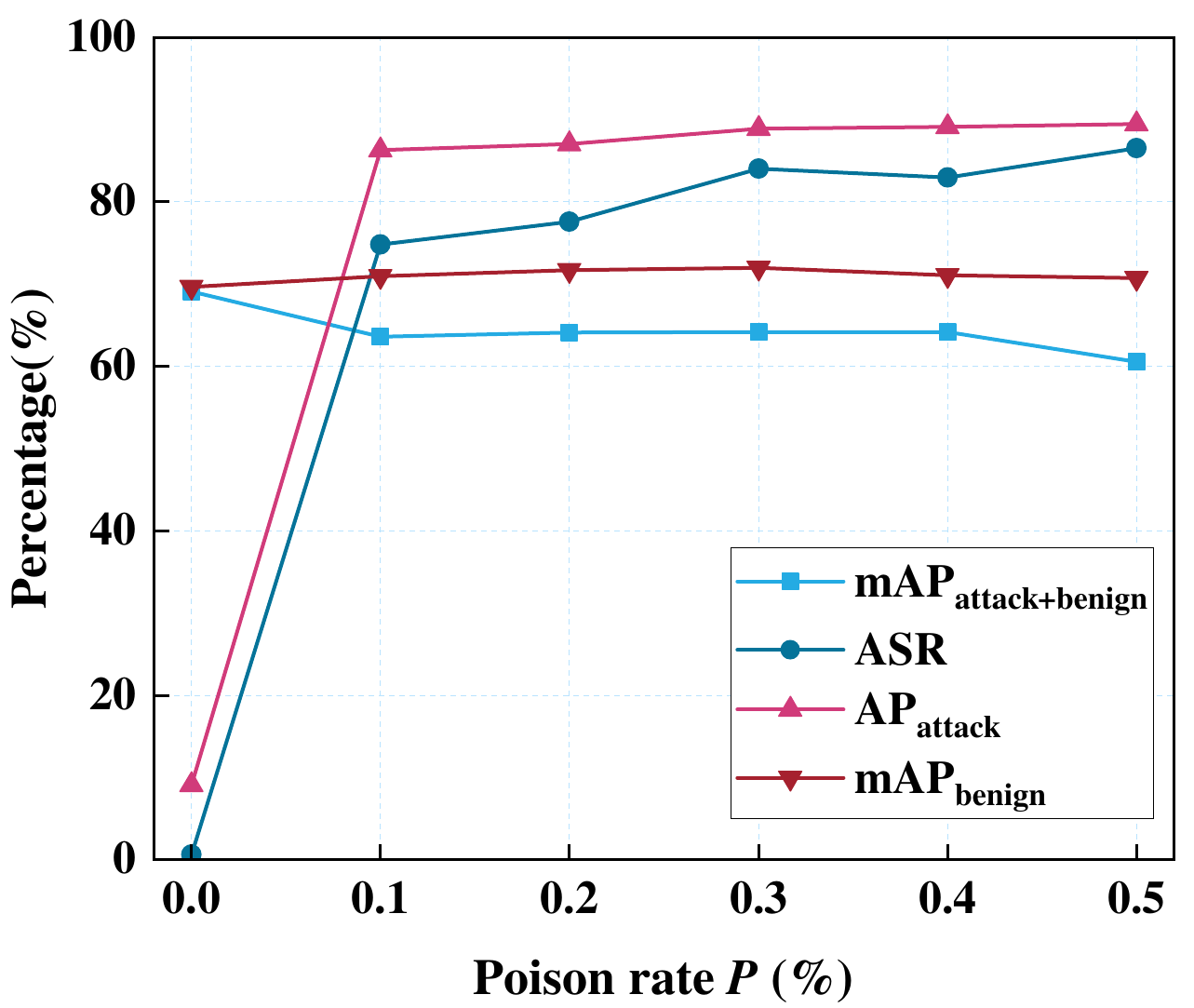}
        \subcaption{}
		\label{fig:3d}
	\end{minipage} 
    \begin{minipage}[c]{0.3\textwidth}
		\includegraphics[width=\textwidth]{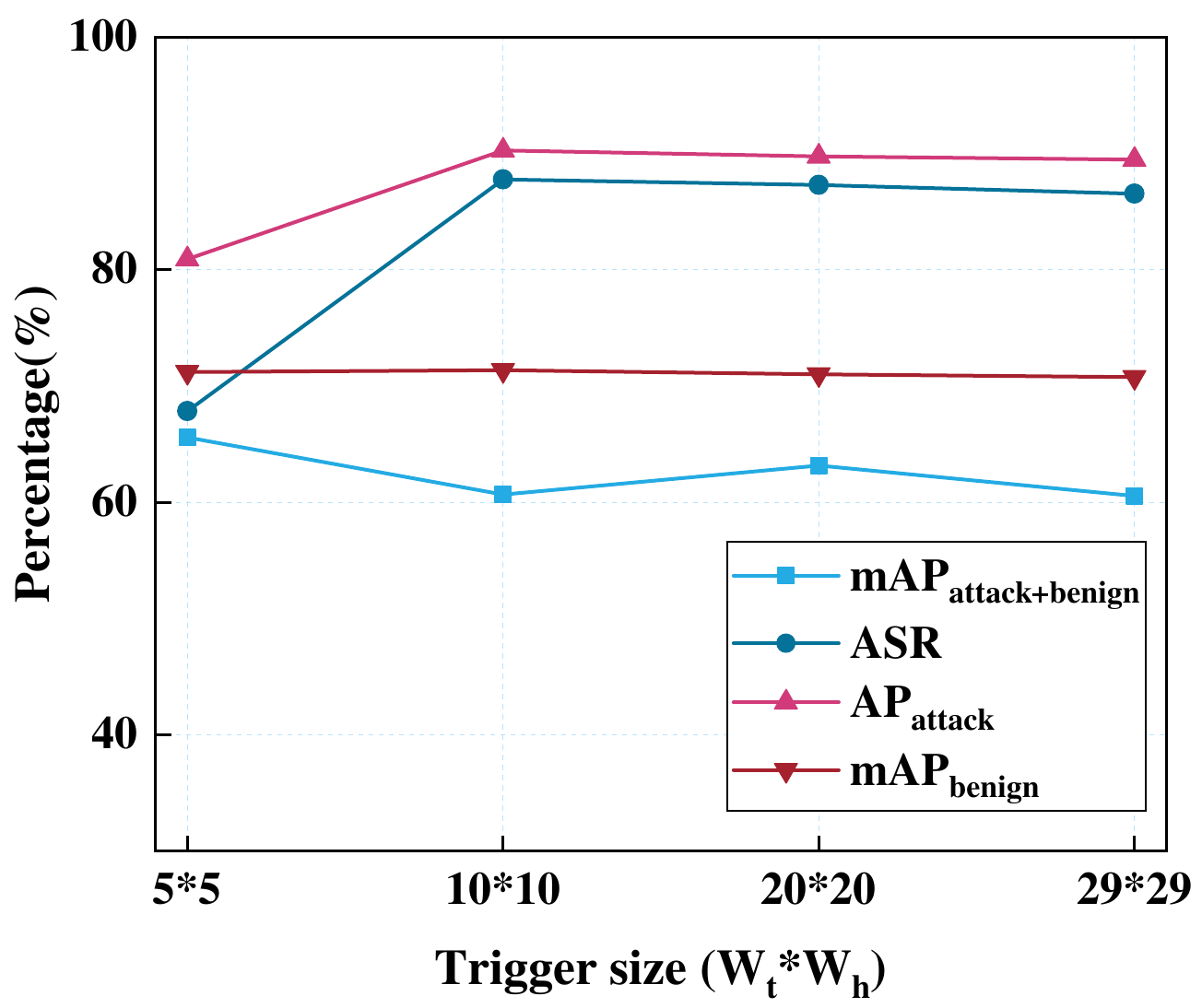}
        \subcaption{}
		\label{fig:3e}
	\end{minipage} 
	\begin{minipage}[c]{0.3\textwidth}
		\includegraphics[width=\textwidth]{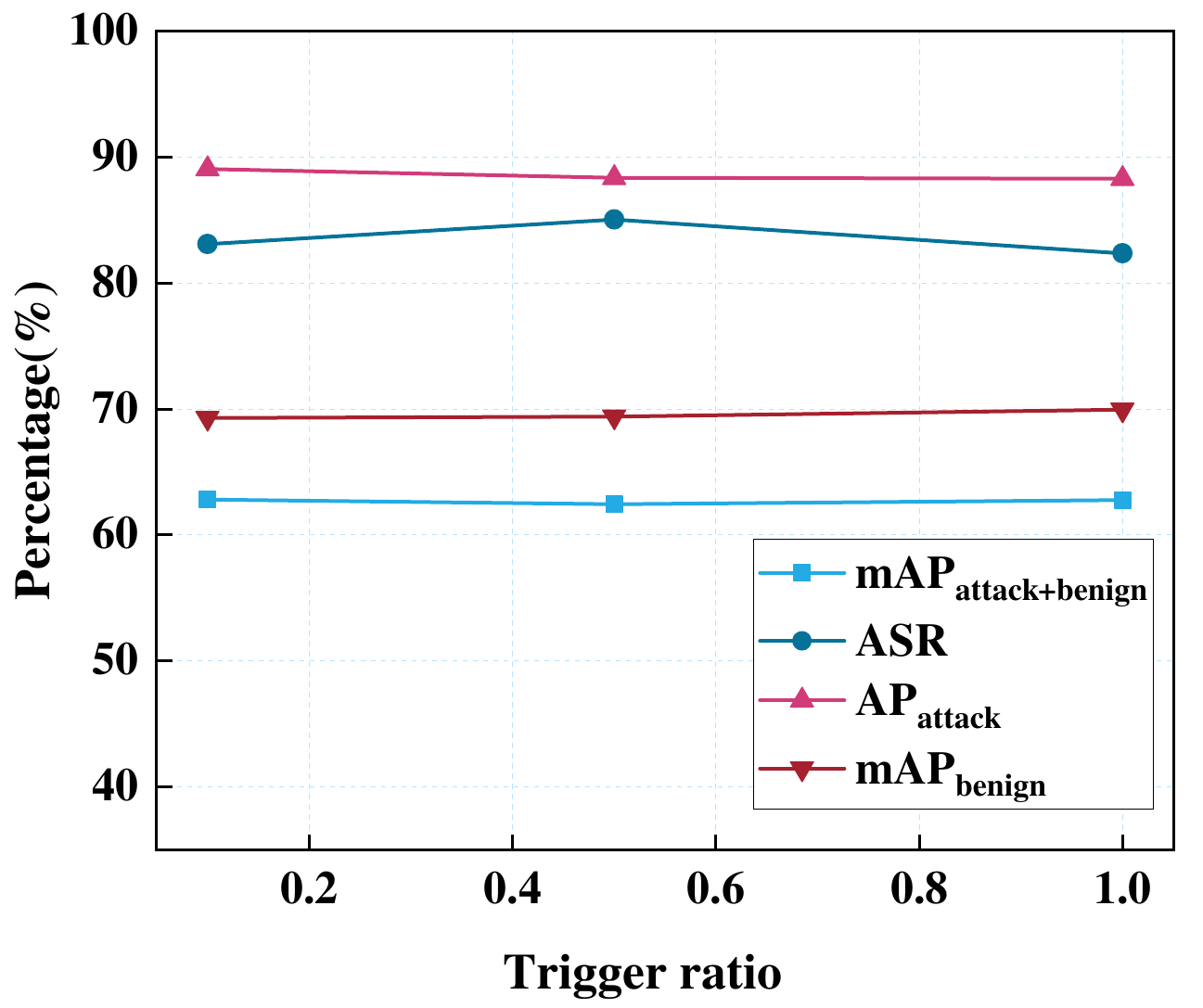}
        \subcaption{}
		\label{fig:3f}
	\end{minipage}
	\caption{
        Impact of poisoning rate, trigger size, and trigger rate on attack effectiveness.
        (a) The impact of poisoning rate on \texttt{NA}.
        (b) The impact of trigger size on \texttt{NA}.
    (c) The impact of trigger ratio on \texttt{NA}.
    (d) The impact of poisoning rate on \texttt{DSBA}.
    (e) The impact of trigger size on \texttt{DSBA}.
    (f) The impact of trigger ratio on \texttt{DSBA}.
    }
	\label{fig:xr}
\end{figure*}

To demonstrate the success of the attack in object detection, we define the Attack Success Rate (ASR) as the extent to which the trigger induces changes in bboxes' categories. An effective $M_{infected}$ should exhibit a high ASR. ASR is calculated as the number of bboxes in $D_{test, backdoored}$ (with confidence $>$ 0.5 and IoU $>$ 0.5) where the predicted results of the attacked category changes to the target category due to the presence of the trigger, divided by the number of bboxes in $D_{test, benign}$ of the attacked categories.

\subsection{Attack Performance}
\label{sec:5.2}
We evaluate $M_{infected}$ using various metrics mentioned in Section\ref{sec:5.1}. For both \texttt{NA} and \texttt{DSBA}, we employ the same settings: a notably low poisoning rate ($P$) of 0.5\%, a trigger size of 29*29, a trigger ratio of 1, and ``person'' as the target category.
We present the detailed results of the two attacks on the shadow dataset with different or similar distributions to the downstream dataset in Table \ref{tab:main_table}. Additionally, we display the evaluation results of $M_{benign}$ in Table \ref{tab: clean model} for comparison with $M_{infected}$.

On the Fast R-CNN model, both \texttt{NA} and \texttt{DSBA} demonstrated strong attack capabilities on the VOC2007 dataset. Notably, \texttt{DSBA} achieved an ASR of 86.55\% on this dataset, significantly higher than \texttt{NA}. As an effective and covert backdoor implantation method in SSL contexts, \texttt{DSBA} exhibits a higher ASR and comparable benign model utility impact. On the MSCOCO dataset, both attacks show high ASR, indicating that \texttt{NA}, even with a very low poisoning rate (e.g., 0.5\%), remains effective in complex datasets. In comparison, \texttt{DSBA}'s Attack Precision ($AP_{attack}$) and ASR improved on MSCOCO, further emphasizing its attack efficacy in complex scenarios.

In most cases, the overall utility loss of both attacks, when comparing $M_{infected}$ to $M_{benign}$, was less than 1\%. \texttt{DSBA} causes a slight decrease in the $mAP_{benign}$ on the $\langle \text{Mask R-CNN, MSCOCO} \rangle$. However, \texttt{DSBA} did not exhibit the expected significant advantage on the more complex MSCOCO dataset. This might be due to \texttt{DSBA}'s involvement with pre-trained encoders and its complex dual-attack strategy, which is more effective in simpler datasets like VOC2007. However, this advantage might be mitigated in MSCOCO, which contains numerous small objects and detailed annotations, due to the inherent noise and diversity of the data. Overall, it is evident that both \texttt{NA} and \texttt{DSBA} are capable of efficient attacks across different datasets and model architectures with minimal impact on model performance. This clearly highlights the potential risks and stealthiness of such attacks in real-world scenarios.

\subsection{Ablation Study}
\label{sec:5.3}

To investigate the different components of the backdoor attacks we introduced, we analyze the influence of poisoning rate ($P$), trigger size $(W_t, H_t)$, trigger ratio $\alpha$, target class $c_t$, and trigger position on both two attacks. Apart from the parameters under study, all other settings remain consistent with those in Section \ref{sec:5.1}. Each ablation study altered a single parameter to observe its effects, and we report both the attack and utility performance in Figure \ref{fig:xr}.


\textbf{The effects of poisoning rates.} As shown in Figure \ref{fig:3a} and Figure \ref{fig:3d}, we explored the impact of poisoning ratios in the downstream fine-tuned dataset on the effectiveness of attacks. We observed that the attack performance of \texttt{NA}, as indicated by metrics such as ASR score and $AP_{attack}$, increases with the rise in poisoning ratio $P$, showing significant overall changes. In contrast, the attack performance of \texttt{DSBA} is relatively less sensitive to changes in the poisoning ratio. Even at lower poisoning levels, \texttt{DSBA} demonstrates substantial effectiveness. For instance, at a poisoning ratio of merely 0.001, \texttt{DSBA}'s $AP_{attack}$ reaches 86.3\%, and its ASR attains 74.8\%. Conversely, \texttt{NA} requires a higher poisoning rate to achieve a similar level of ASR. Regarding the model's practical performance, we noted that parameters such as $mAP_{benign}$ slightly decrease as the poisoning ratio increases, with the reduction in \texttt{NA} being slightly greater than that in \texttt{DSBA}. This indicates that both attacks can maintain high attack performance while ensuring the model's utility.


\textbf{The effects of trigger size.} We also demonstrated the impact of trigger size on attack performance, as shown in Figures \ref{fig:3b} and Figure \ref{fig:3e}. We found that larger trigger sizes enhance attack effectiveness. The success rate of \texttt{NA} increases with the size of the trigger. A small trigger size (such as $5*5$) significantly reduces attack performance, whereas a size of $10*10$ ensures relatively stable attack outcomes. In contrast, \texttt{DSBA} exhibits low sensitivity to trigger size, maintains a steady ASR across different trigger sizes, and achieves a high ASR of 85.48\% with a trigger size of $10*10$. Notably, \texttt{DSBA} maintains commendable attack performance even with a minimal trigger size of $5*5$, achieving an ASR of 67.83\% and an $mAP_{benign}$ of 71.18\%. This contributes to making the trigger more difficult to detect, thereby enhancing the stealth of the attack.


\textbf{The effects of trigger ratio.} Furthermore, our study investigated the impact of trigger rates $\alpha$ on attack outcomes, as illustrated in Figures \ref{fig:3c} and \ref{fig:3f}. We discovered that higher trigger rates, denoted by $\alpha$, have a minimal effect on metrics such as ASR and $mAP_{benign}$ across both attacks. In most scenarios, regardless of the trigger ratio settings, the ASR for \texttt{NA} remains close to 73\%, while for \texttt{DSBA}, it hovers around 85\%, with both maintaining a $mAP_{benign}$ similar to that of a clean model (approximately 70\%). This indicates that attackers could employ minimal trigger ratios (e.g., $\alpha=0.1$) to render the trigger virtually invisible in images, suggesting a high potential for stealth in attack implementation.

\textbf{The effects of target class and trigger position.} 
To assess the impact of target class variation, we modified the target class from ``dog'' to ``sheep,'' a category with fewer instances. According to the results presented in Table \ref{subtab:sheep_class}, the performance of the \texttt{NA} and \texttt{DSBA} methods is not adversely affected by the reduced number of target class objects. Furthermore, to demonstrate that the placement of the trigger does not influence the outcome of the attack, we altered the position of the trigger to random locations within the images of the poisoned dataset, rather than centering it within the bbox. The findings, as shown in Table \ref{subtab:random_location}, indicate that this alteration does not affect the efficacy of the attack, paralleling the results observed in Table \ref{tab:main_table}.

\begin{table}[t]
    \renewcommand{\arraystretch}{1.2}
    \caption{
    Attack Performance ($\%$) When (a) Target Class is Changed to "Sheep" Class and (b) Trigger's Locations are Changed to Random Locations. “$\uparrow$”/“$\downarrow$”/“\text{-}” follow definitions in Table \ref{tab:main_table}.} 
    \label{tab:sheep}
    \centering
    \begin{subtable}{\linewidth}
        \centering
        \caption{Target class t = “sheep” class.}
        \label{subtab:sheep_class}
        \resizebox{0.9\linewidth}{!}{
        \begin{tabular}{c|cc}
           \toprule[1.2pt]
           Attack Method & \texttt{NA} & \texttt{DSBA} \\
           \midrule
          $mAP_{benign}$ \text{-} & 69.62 & 71.14\\
          $AP_{benign}$ \text{-} & 50.98 & 48.61 \\
          $mAP_{attack}$ $\uparrow$ & 63.89 & 68.20\\
          $AP_{attack}$ $\uparrow$ & 76.65 & 89.53 \\
          $mAP_{attack+benign}$ \text{-} & 61.62 & 64.46\\
          $AP_{attack+benign}$ $\downarrow$ & 0.30 & 0.23 \\
          $mAP_{part\_attack+benign}$ \text{-} & 60.08 & 62.83 \\
          \midrule
          $ASR$ $\uparrow$ & 74.52& 87.04\\
           \bottomrule[1.2pt]
        \end{tabular}
        }
    \end{subtable}

    \vspace{0.7cm}
    
    \begin{subtable}{\linewidth}
        \centering
        \caption{Random triggers’ locations.}
        \label{subtab:random_location}
        \resizebox{0.9\linewidth}{!}{
        \begin{tabular}{c|cc}
           \toprule[1.2pt]
           Attack Method & \texttt{NA} & \texttt{DSBA} \\
           \midrule
          $mAP_{benign}$ \text{-} & 69.97 & 72.44\\
          $AP_{benign}$ \text{-} & 69.49 & 72.25 \\
          $mAP_{attack}$ $\uparrow$ & 61.52 & 65.88\\
          $AP_{attack}$ $\uparrow$ & 79.74 & 88.05 \\
          $mAP_{attack+benign}$ \text{-} & 62.52 & 64.92\\
          $AP_{attack+benign}$ $\downarrow$ & 1.67 & 1.74 \\
          $mAP_{part\_attack+benign}$ \text{-} & 60.34 & 62.18 \\
          \midrule
          $ASR$ $\uparrow$ & 71.44 & 82.12 \\
           \bottomrule[1.2pt]
        \end{tabular}
        }
    \end{subtable}
\end{table}

\begin{figure}[htbp]
	\centering
	\begin{minipage}[c]{0.35\textwidth}
		\includegraphics[width=\textwidth]{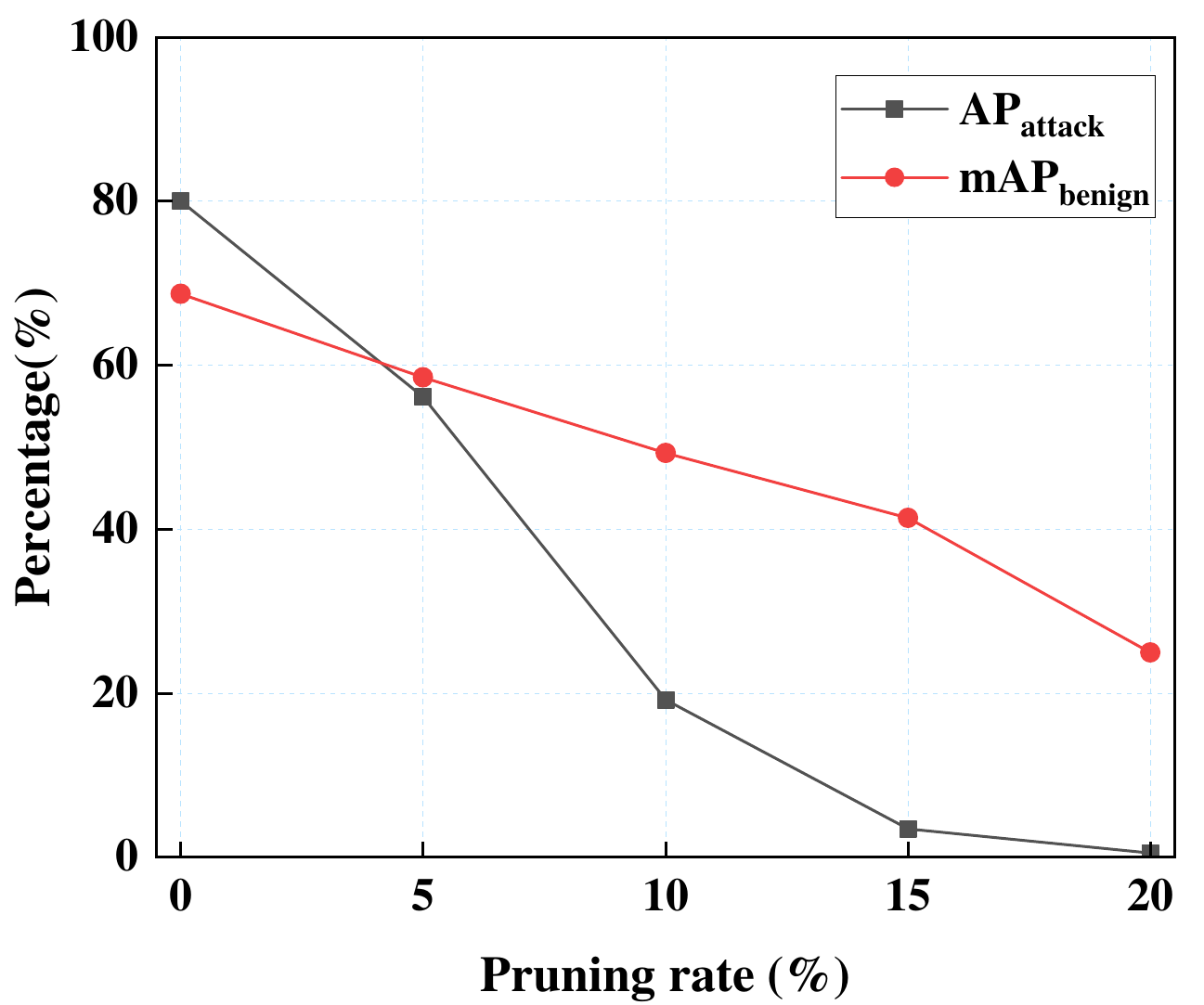}
        \subcaption{}
		\label{fig:4a}
	\end{minipage}
	\begin{minipage}[c]{0.35\textwidth}
		\includegraphics[width=\textwidth]{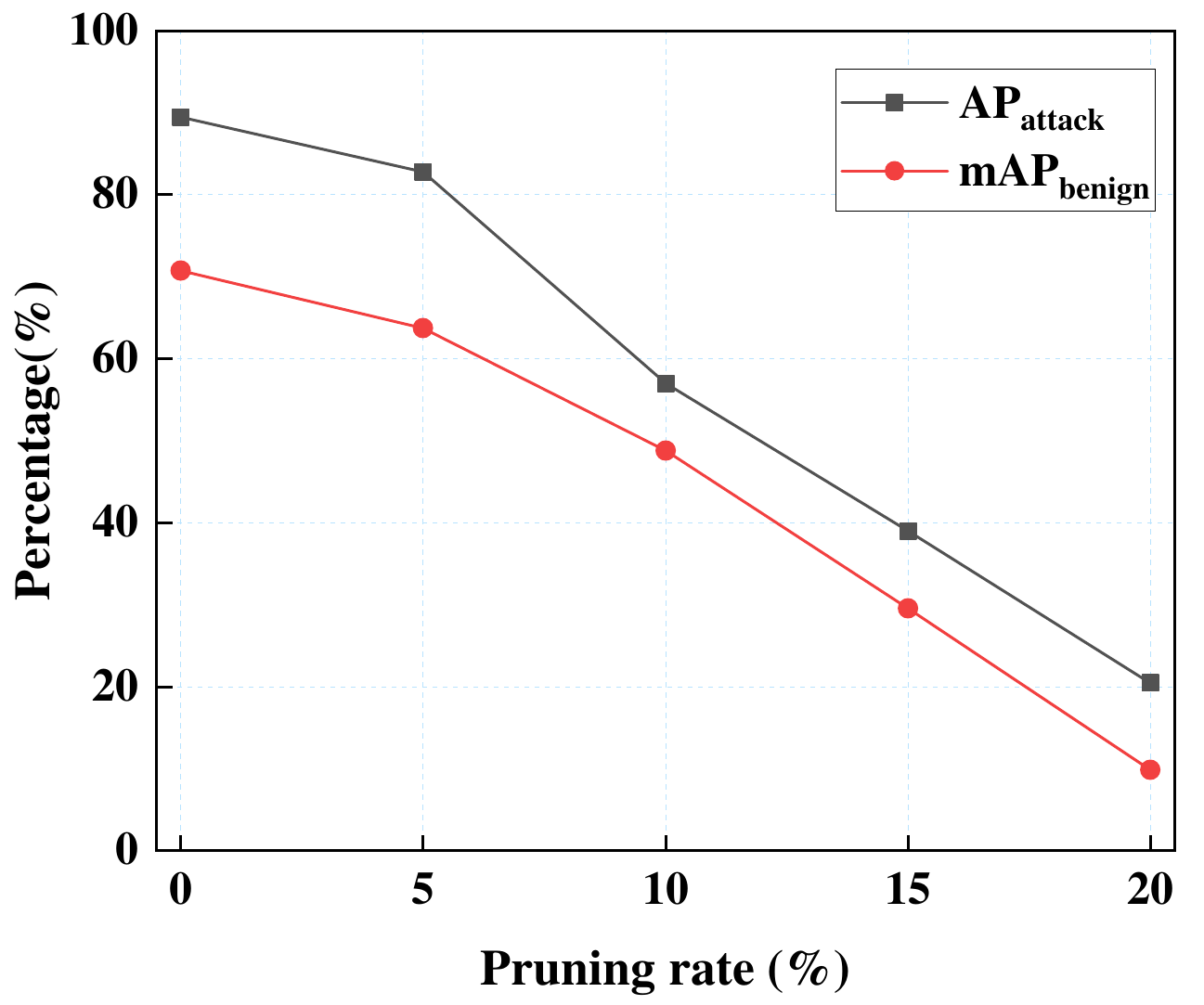}
        \subcaption{}
		\label{fig:4b}
	\end{minipage}

	\caption{
        The resistance to model pruning.
        (a) The effect of \texttt{NA} against model pruning.
        (b) The effect of \texttt{DSBA} against model pruning.
    }
	\label{fig:pruning}
\end{figure}

Overall, \texttt{DSBA} demonstrates greater flexibility and robustness in executing effective backdoor attacks, offering superior performance and versatility across various aspects compared to \texttt{NA}, which is effective under specific conditions. Moreover, both attacks minimally impact model utility (as indicated by metrics such as $mAP_{benign}$), suggesting they can successfully remain undetected within the model.

\subsection{The Resistance to Model Pruning}
\label{sec:5}

Although many defense or detection methods have achieved significant success in addressing backdoor attacks on image classification, these methods face numerous challenges when applied to object detection tasks. First, methods~\cite{wang2019neural,shen2021backdoor} that predict the distribution of backdoor triggers through generative modeling or neuron reverse engineering typically assume that the model is a simple neural network. In contrast, object detection models consist of multiple complex components, making it difficult to apply these methods directly. Furthermore, the output of image classification models is a single predicted category, while object detection models output multiple objects along with their positions and categories. This diversity in output further complicates the detection of backdoor triggers.

In the SSL scenario, there are currently no specific defense measures against backdoor attacks on object detection models. Therefore, we consider using general backdoor defense strategies to evaluate the robustness of our proposed attack. Model pruning~\cite{liu2018fine,wu2021adversarial} is a typical backdoor defense method that can be widely applied to various tasks. It reduces the impact of backdoor triggers by removing neurons with low activation rates in benign datasets. In this section, we will evaluate whether our attack can withstand this potential defense. Specifically, after model training is complete, we perform model pruning during the testing phase. Using 40\% of the benign test samples, we prune neurons with the lowest activation values by comparing the model's performance on the backdoor test dataset and the clean dataset. As shown in Figure \ref{fig:pruning}, as the pruning rate increases, the $AP_{attack}$ values for both attacks decrease, but the $mAP$ values for benign samples also decrease rather than increase. This indicates that model pruning has limited effectiveness in object detection tasks. Object detection models need to retain more neurons to handle complex detection tasks, and pruning can lead to a significant decline in model performance (e.g., $mAP$). In summary, our method can withstand model pruning.

\section{Related Work}
\subsection{Backdoor Attacks}

Backdoor attacks are an emerging topic with potential development prospects in the field of network security, incorporating specially designed triggers into inputs. This compels the model embedded with a backdoor to classify it into a predefined target category with high confidence. Importantly, the backdoor model can still function normally and produce almost the same accuracy as an uninfected model even after the removal of the trigger from the inputs.

In the field of supervised learning, backdoor attacks targeting image classifiers have been broadly explored by researchers. These attacks are designed to misclassify images containing triggers chosen by the attacker, typically by contaminating the training dataset. Such forms of attacks were initially studied in the context of supervised learning~\cite{gu2017badnets,liu2018trojaning}. More advanced and difficult-to-detect methods of attack have been proposed in~\cite{quiring2020backdooring,saha2020hidden,zhang2022poison,wang2022invisible}, exploring covert attacks under various scenarios. For instance, Quiring et al.~\cite{quiring2020backdooring} investigated attacks targeting image scaling, while Saha et al.~\cite{saha2020hidden} assumed that attackers are aware of the model structure. In~\cite{zhang2022poison}, researchers utilized image structures to create covert trigger areas and embedded them into images using deep injection networks, achieving covert attacks. Similarly, a method involving the conversion from RGB to YUV channels as well as DCT transformations for embedding triggers was proposed in~\cite{wang2022invisible}.

The area of SSL offers a unique perspective for backdoor attacks. Recent studies, such as those by Saha et al.~\cite{saha2022backdoor}, have demonstrated the embedding of backdoors into the unlabeled datasets of SSL without damaging the training phases of image encoders and downstream classifiers. Building upon this, Li et al.~\cite{li2023embarrassingly} defined trigger patterns as specific disturbances in the frequency domain. However, these attacks have shown shortcomings in terms of attack efficacy and model performance. Other concurrent works~\cite{xue2022estas,badencoder} have attempted to improve the ASR. Specifically, BadEncoder~\cite{badencoder} focuses on executing precise backdoor attacks in SSL encoders by compromising encoder training, assuming access to a clean, pre-trained encoder. This approach has been proven to achieve more than 98\% ASR on the ImageNet-100 dataset. Research into backdoor attacks and their defense/detection methods has been extensively explored across multiple fields, including image recognition, video recognition, natural language processing (e.g., sentiment classification, toxicity detection, spam detection), and federated learning. However, existing backdoor attacks on self-supervised pre-trained models have largely focused on image classification tasks, with such attacks on object detection tasks yet to be explored.

\subsection{Object Detection Using SSL}

Significant progress has been made in the field of SSL in recent years. However, most methods~\cite{he2020momentum,grill2020bootstrap} are primarily designed for image classification, though they also demonstrate transferability to object detection tasks. For instance, the representations learned by MoCo~\cite{he2020momentum} can be effectively transferred to downstream tasks. Compared to supervised pretraining on ImageNet, MoCo has shown superior transfer capabilities in various object detection tasks. 

Although direct applications of SSL in object detection have been less common, recent works like DenseCL~\cite{wang2021dense}, InsLoc~\cite{yang2021instance}, and PatchReID~\cite{Ding2021UnsupervisedPF} have begun to explore SSL pretraining tasks specifically designed for object detection. DenseCL enhances local feature learning by conducting dense contrastive learning at the pixel level. InsLoc combines instance-level contrastive learning, resulting in superior instance-level feature representations. PatchReID employs a local region contrast strategy, further improving the model's performance in fine-grained feature learning for object detection. These methods leverage contrastive learning to improve object detection performance, indicating that SSL techniques are gradually covering a broader range of tasks within the computer vision field.

\section{Conclusion}

In this work, we investigate, for the first time, backdoor attacks on downstream tasks of object detection in the SSL scenario, elucidating the potential backdoor threats faced by object detection in this context. In the SSL-OTA, we propose a simple yet effective backdoor attack strategy (\texttt{NA}) and further introduce an enhanced attack strategy (\texttt{DSBA}). We define appropriate metrics for evaluating attack performance, demonstrating the effectiveness and utility of both attack strategies. Additionally, we delve into an in-depth exploration of the impact of various parameters and triggering factors on the efficacy of the attacks. Our approach serves as a valuable tool for assessing the backdoor robustness of object detectors, contributing to the design of more secure models.

\bibliographystyle{IEEEbib}
\bibliography{icdm2024template}
\vspace{12pt}

\end{document}